**Surface roughness influence on the quality factor of high frequency nanoresonators**

G. Palasantzas [a]

Zernike Institute of Advanced Materials, University of Groningen, Nijneborgh 4, 9747 AG Groningen, The Netherlands

**Abstract**

Surface roughness influences significantly the quality factor of high frequency nanoresonators for large frequency - relaxation times ($\omega\tau>1$) within the non-Newtonian regime, where a purely elastic dynamics develops. It is shown that the influence of sort wavelength roughness, which is expressed by the roughness exponent $H$ for the case of self-affine roughness, plays significant role in comparison with the effect of the long wavelength roughness parameters such as the rms roughness amplitude and the lateral roughness correlation length. Therefore, the surface morphology can play important role in designing high-frequency resonators operating within the non-Newtonian regime.

*Pacs numbers:* 85.85.+j, 73.50.Td, 68.55.-a, 74.62.Fj

---

[a] Corresponding author: G.Palasantzas@rug.nl

Nanoelectromechanical systems (NEMS) are important devices that combine the advantages of mechanical systems (e.g., as sensor systems and robustness to electrical shocks) with the speed and large scale integration of silicon based microelectronics [1-11]. Nowadays, there is an intense effort to develop NEMS operating in dense gaseous and liquid environments for applications in nanofluidics and bioengineering, where high-frequency oscillating microflows can take place violating the Newtonian approximation [8-11]. Moreover, as the resonator size is reduced the surface to volume ratio increases, which makes the nanoresonators susceptible to a variety of surface related dissipation mechanisms.

Indeed, the central theme that underlies a large part of research in NEMS is the achievement of ultrahigh quality factors $Q$ (stored to dissipated energy within an oscillation cycle), which is affected strongly by the surface condition (oxides, defects, and roughness etc.). In a variety of studies it has been shown that surface roughness decreases the quality factor for operation at low pressure environment [12, 13]. Studies for SiC/Si NEMS indicted that devices operational in the UHF/microwave regime had a low surface roughness (~ 2 nm), while devices with rougher surfaces (~ 7 nm) were operational only up to the VHF range [12]. For Si nanowires it was shown that the quality factor was decreased by increasing the surface to volume ratio [13]. Recently, within the molecular regime (molecule mean free path larger than the lateral nanoresonator dimensions), it was shown that random surface roughness can decrease the quality factor and dynamic range, and increase the limit to mass sensitivity [14-16].

Furthermore, under operation in dense fluid a high resonance-frequency nanomechanical resonator generates a rapidly oscillating flow in the surrounding fluid environment [10, 11]. Over a broad frequency ($\omega$) and pressure ($P$) range explored, it was observed a sign of a transition from Newtonian to non-Newtonian flow at $\omega\tau\approx1$ with $\tau$ a fluid relaxation time [10, 11]. The latter was confirmed experimentally in close quantitative agreement with theory, which predicts purely elastic fluid response as $\omega\tau\rightarrow\infty$ [10, 11]. The Newtonian approximation (the basis for the Navier-Stokes equations) breaks down when the particulate nature of the fluid becomes significant to the flow. It is common to consider application or not of the Newtonian approximation by comparing the mean free path $\lambda_D$ in the medium to an ill-defined characteristic length $L$ or using the Knudsen number $Kn = \lambda_D/L$. For oscillatory flow, another defining parameter is the Weissenberg number $Wi=\tau/T$, which compares the characteristic time scale $T$ of the flow with the relaxation time $\tau$ in the medium. Varying the ratio $\tau/T$ can lead to drastic changes in the nature of the flow around the resonator.

Therefore, in this work we will extend the previous studies in a dense fluid environment for high frequency nanoresonators to the case with rough surfaces. Focus will be given in the regime where the Newtonian to non-Newtonian flow takes place ($\omega\tau\approx1$) [10, 11]. For simplicity we will consider for the roughness description the case of random self-affine roughness, which is observed in various surface engineering processes. The latter is characterized by the roughness amplitude $w$, the lateral roughness correlation length $\xi$, and the roughness exponent $0<H<1$ that characterize the degree of surface irregularity at short length scales ($<\xi$).

In the following, the resonator motion is described the one-dimensional damped harmonic oscillator $\ddot{x}+\gamma\dot{x}+\omega^2 x = F/m_{res}$ [3] with $F$ the force acting on an effective oscillating mass $m_{res}$. The quality factor $Q_{f,rough}$ is related to fluidic dissipation $\gamma$ by means of the relation [10, 11]

$$\gamma = (\omega/Q_{f,rough})(m_{res}/A_{rough})$$

$$\gamma = (1+\omega^2\tau^2)^{-3/4}(\omega\mu\rho_f/2)^{1/2}$$
$$x[(1+\omega\tau)\cos(\tan^{-1}(\omega\tau)/2)-(1-\omega\tau)\sin(\tan^{-1}(\omega\tau)/2)]$$
(1)

with $\mu$ the fluid viscosity, $\rho_f$ the mass density of the surrounding fluid, and $A_{rough}$ the area of the resonator that is assumed randomly rough. The dependence on the ratio $m_{res}/A_{rough}$ stems from the fact that dissipation into the fluid is proportional to the effective surface area $A_{rough}$, while the stored energy in the resonator is proportional to the effective oscillating mass $m_{res}$ [10, 11].

Furthermore, we assume for the roughness profile a single valued random fluctuation $h(r)$ of the in-plane position $r=(x,y)$. For a Gaussian height distribution the surface area $A_{rough}$ is given by $A_{rough}/A_{flat}=\int_0^{+\infty}du\sqrt{1+\rho^2 u}e^{-u}$ [17] with $A_{flat}$ the average flat surface area, and $\rho=<\sqrt{<(\nabla h)^2>}$ the average local surface slope. Substitution in Eq. (1) yields

$$Q_{f,rough}=(\omega/\gamma)(m_{res}/A_{flat})(\int_0^{+\infty}du\sqrt{1+\rho^2 u}e^{-u})^{-1}.$$
(2)

If we substitute in $\rho$ the Fourier transform of the surface height $h(q)=(2\pi)^{-2}\int h(r)e^{-iq\cdot r}d^2r$ and assume translation invariance so that $<h(q)h(q')>=\delta^2(q'+q)C(q)$, we obtain $\rho^2=\int_{0\leq q\leq Q_c} q^2 <|h(q)|^2> d^2q$ where $Q_c=\pi/a_o$ with $a_o$ a lower roughness cut-off of the order of atomic dimensions. $<|h(q)|^2>$ is the Fourier transform of the height correlation function $C(r)=<h(r)h(O)>$, and its knowledge is necessary for the calculation of $Q_{f,rough}$. Indeed, a wide variety of random surfaces possess the so-called self-affine roughness, with a spectrum that scales as $<|h(q)|^2> \propto q^{-2-2H}$ if $q\xi>>1$, and $<|h(q)|^2> \propto const$ if $q\xi<<1$ [18]. This is satisfied by the analytic roughness model $<|h(q)|^2>=(w^2\xi^2/2\pi)/(1+aq^2\xi^2)^{(1+H)}$ with $a=(1/2H)[1-(1+aQ_c^2\xi^2)^{-H}]$ if $0\leq H\leq 1$ [19]. Small values of $H$ (~0) characterize jagged or irregular surfaces; while large values of $H$ (~1) surfaces with smooth hills-valleys (inset, Fig. 1) [18, 19]. Substitution of $<|h(q)|^2>$ yields the local slope the analytic form $\rho=(w/\sqrt{2}\xi a)[(1-H)^{-1}[(1+aQ_c^2\xi^2)^{1-H}-1]-2a]^{1/2}$ [20].

In the following, the calculations of the quality factor were performed for roughness amplitudes observed in real resonators in the range $w\sim 2$-$8$ nm [8]. Figure 1 shows the quality factor for various roughness exponents $H$. It becomes evident that the rougher the surface is at short wavelengths ($<\xi$) as described by the exponent $H$ (inset, Fig. 1), the more is magnified its effect on the quality factor for relatively large values of $\omega\tau$ within the non-Newtonian regime ($\omega\tau>1$), where the flow around the resonator shows an elastic response. Similar is the situation with decreasing correlation length $\xi$ as it is shown in Fig. 2, and corresponding to surface roughening at long wavelengths (decreasing $\xi$ and/or increasing *w*). Comparisons of Figs. 1 and 2 shows that the variation of the quality factor is more drastic with decreasing roughness exponent *H* in the range of values [0,1]. For weak roughness or equivalently small local surface slopes $\rho<<1$ (so that $A_{rough}/A_{flat}\approx 1+\rho^2/2$) the $Q_{f,rough}$ is given by the analytic form

$$Q_{f,rough} \approx (\omega/\gamma)(m_{res}/A_{flat})(1-(w^2/4\xi^2 a^2)\{(1-H)^{-1}[(1+aQ_c^2\xi^2)^{1-H}-1]-2a\}).$$

If we consider the derivative of the quality factor with respect to $\omega\tau$, we obtain, as Fig. 3 indicates, a minimum approximately for $\omega\tau\approx 1/2$. For $\omega\tau<<1$ the derivative of the quality factor increases rapidly with decreasing $\omega\tau$, while with increasing roughness the minimum drastically attenuates. On the other hand, for large $\omega\tau$ (>>1) the derivative of the quality factor appears almost constant. In any case, the minimum indicates that the transition from viscous to elastic regime occurs at lower frequencies $\omega$ by assuming a fixed relaxation time $\tau$.

Finally, we express the quality factor as a function of the fluid pressure *P* through the experimentally confirmed relation $\tau\approx 1850/P$ [10]. The is shown in Fig. 4 for different roughness exponents *H*, and lateral correlation lengths $\xi$ (inset). In fact, $Q_{f,rough}$ decreases with increasing pressure or decreasing relaxation time assuming fixed oscillation frequency $\omega$, but at rate that depends on the particular surface roughness parameters. Comparing the schematics in Fig. 4 it is shown again that the roughness exponent *H* plays dominant role for low pressures or large relaxation times. Despite that $Q_{f,rough}$ decreases faster for larger roughness exponents *H* and/or larger correlation lengths $\xi$ (equivalently smoother surfaces), a weak minimum is observed to occur with increasing pressure *P*. The minimum is more pronounced with decreasing surface roughness.

It should be pointed out that Eq. (1) ignores vortex and shedding formation, which can be generated by surface roughness elements increasing the dissipation. These effects are well-known from molecular dynamics simulations [21], but it is a difficult problem to deal from the purely theoretical point of view. Moreover, with triangular rough surfaces (riblets), if vortices are formed with a radius approximately the distance of riblets then the vortices contact only the peaks of the riblets and friction is reduced resulting in faster fluid flow [22]. At any rate if we assume for simplicity that the resonator surface roughness is represented by triangular roughness elements of height $w$ (~2-8 nm), then if $w<<\delta$ with $\delta = \sqrt{2\mu/\rho_f \omega}$ the boundary layer thickness the influence of turbulent behaviour is expected to be weak. Indeed, this condition is satisfied since $\delta \approx 1-5$ $\mu m$ for the conditions in (and in the present study) [10]. Note that the boundary layer is also influenced by surface roughness [23]. In any case, dissipation associated with vortex formation and shedding will be considered in more detail in our future studies as a function of increasing surface roughness.

In conclusion, we showed that the presence of surface roughness on high frequency nanoresonators influences their frequency dependent quality factor especially for frequency-relaxation time $\omega\tau>1$. The influence of sort wavelength roughness details though the roughness exponent $H$ appears to play significant role. Therefore, our results will possibly play significant role in designing resonators operating under non-Newtonian conditions.

**Figure Captions**

**Figure 1** Quality factor vs. $\omega\tau$ for $C_f = (m_{res}/A_{flat})(1/\tau\mu\rho_f/2)^{1/2}$, $w=5$ nm, $\xi=100$ nm, $a_o=0.3$ nm, and $H$ as indicated. The inset shows examples of roughness profiles wit the same roughness amplitude w but different roughness exponents $H$.

**Figure 2** Quality factor vs. $\omega\tau$ for $C_f = (m_{res}/A_{flat})(1/\tau\mu\rho_f/2)^{1/2}$, $w=5$ nm, $a_o=0.3$ nm, correlation length $\xi$ as indicated, and $H=0.5$.

**Figure 3** Derivative of the quality factor (with respect to $\omega\tau$) vs. $\omega\tau$ for $C_f = (m_{res}/A_{flat})(1/\tau\mu\rho_f/2)^{1/2}$, $w=5$ nm, $\xi=100$ nm, $a_o=0.3$ nm, and $H$ as indicated.

**Figure 4** Quality factor vs. pressure P for $C_f = (m_{res}/A_{flat})(1/\mu\rho_f/2)^{1/2}$, $\omega=1$ MHz, $\tau=1850/P$ (in ns), for $w=5$ nm, $\xi=100$ nm, $a_o=0.3$ nm, and $H$ as indicated. The inset shows the quality factor vs. pressure P for $C_f = (m_{res}/A_{flat})(1/\mu\rho_f/2)^{1/2}$, $\omega=1$ MHz, $\tau=1850/P$ (in ns), for $w=5$ nm, $H=0.5$, $a_o=0.3$ nm, and $\xi$ as indicated.

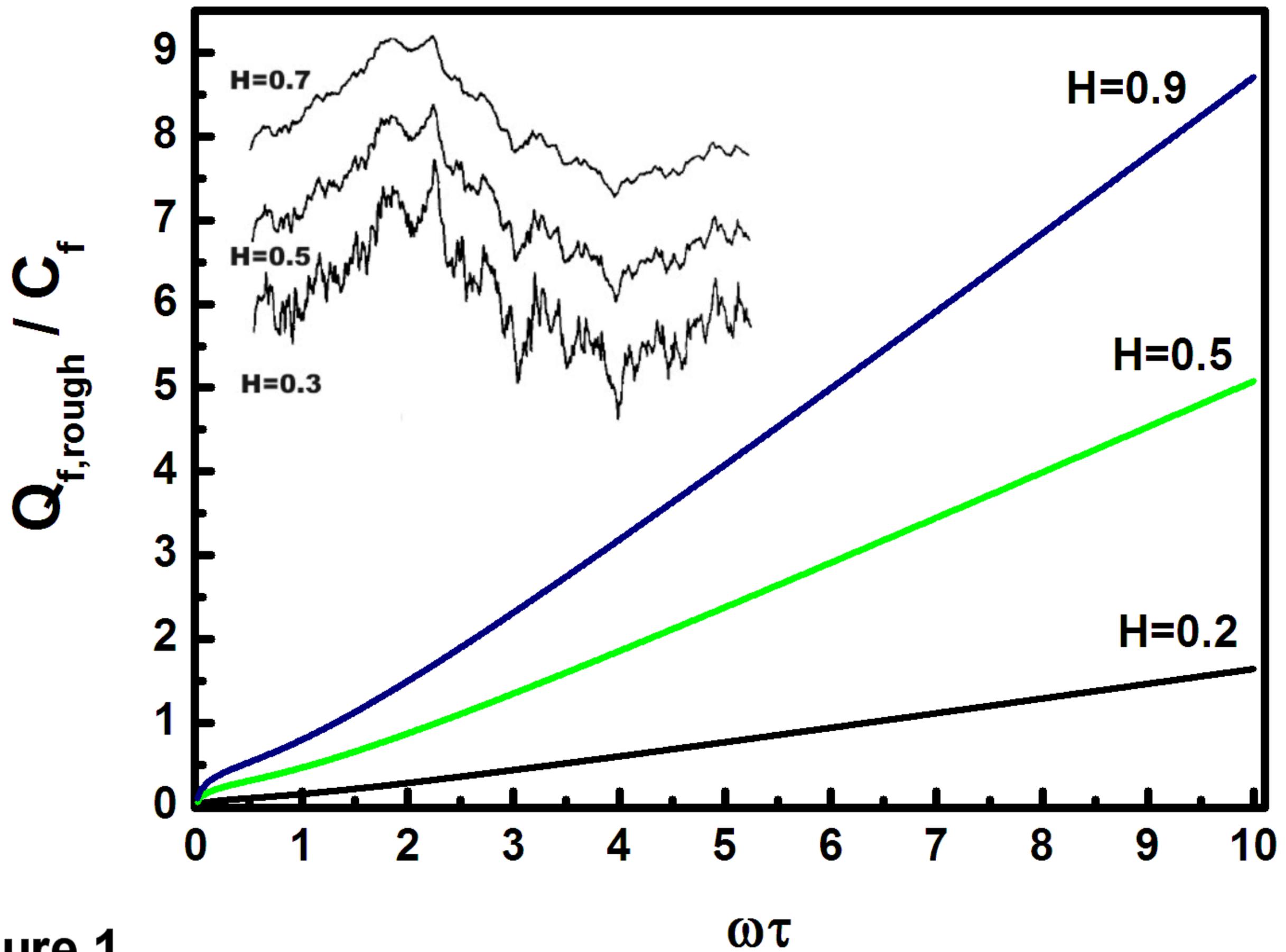

Figure 1

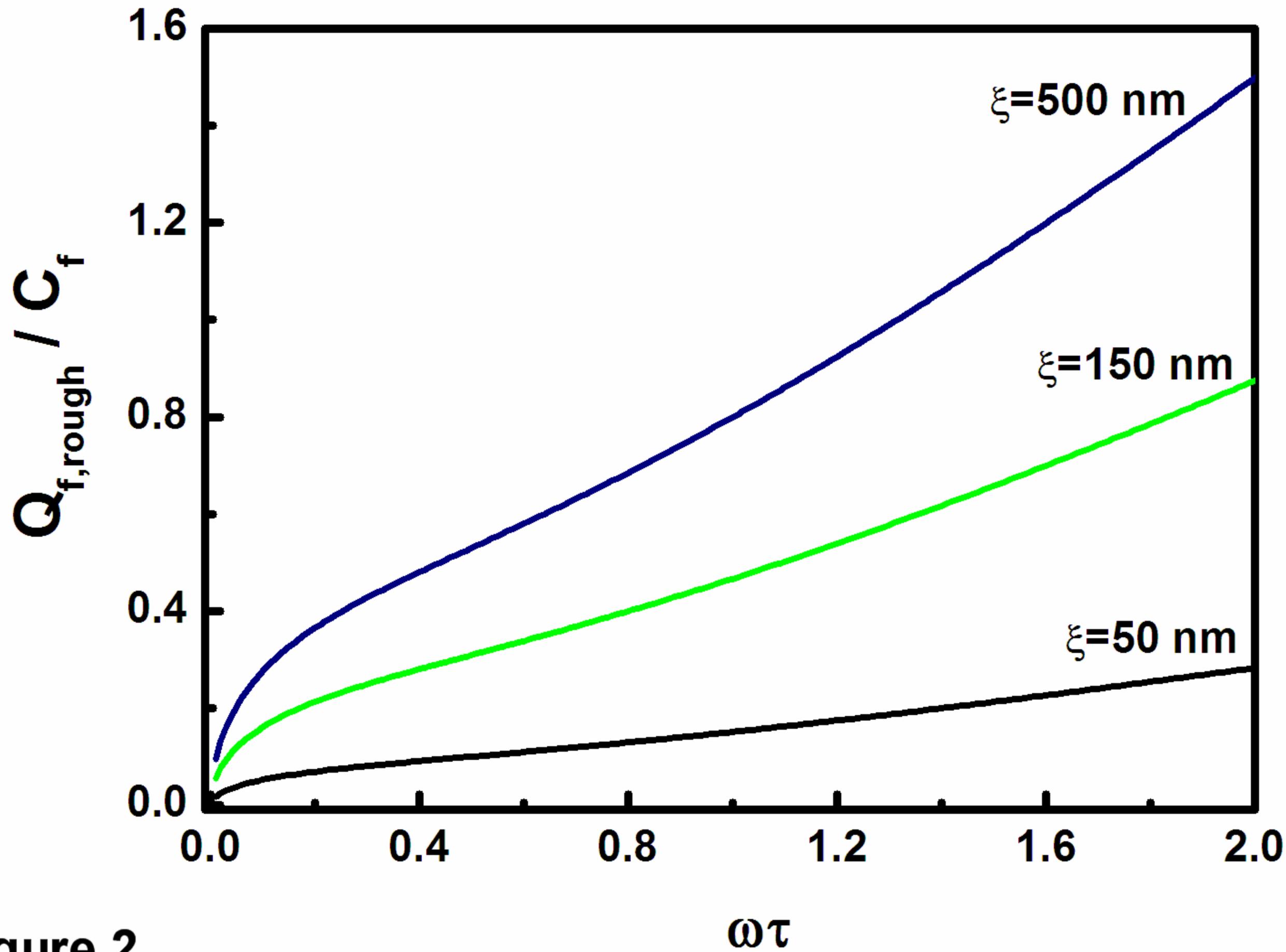

**Figure 2**

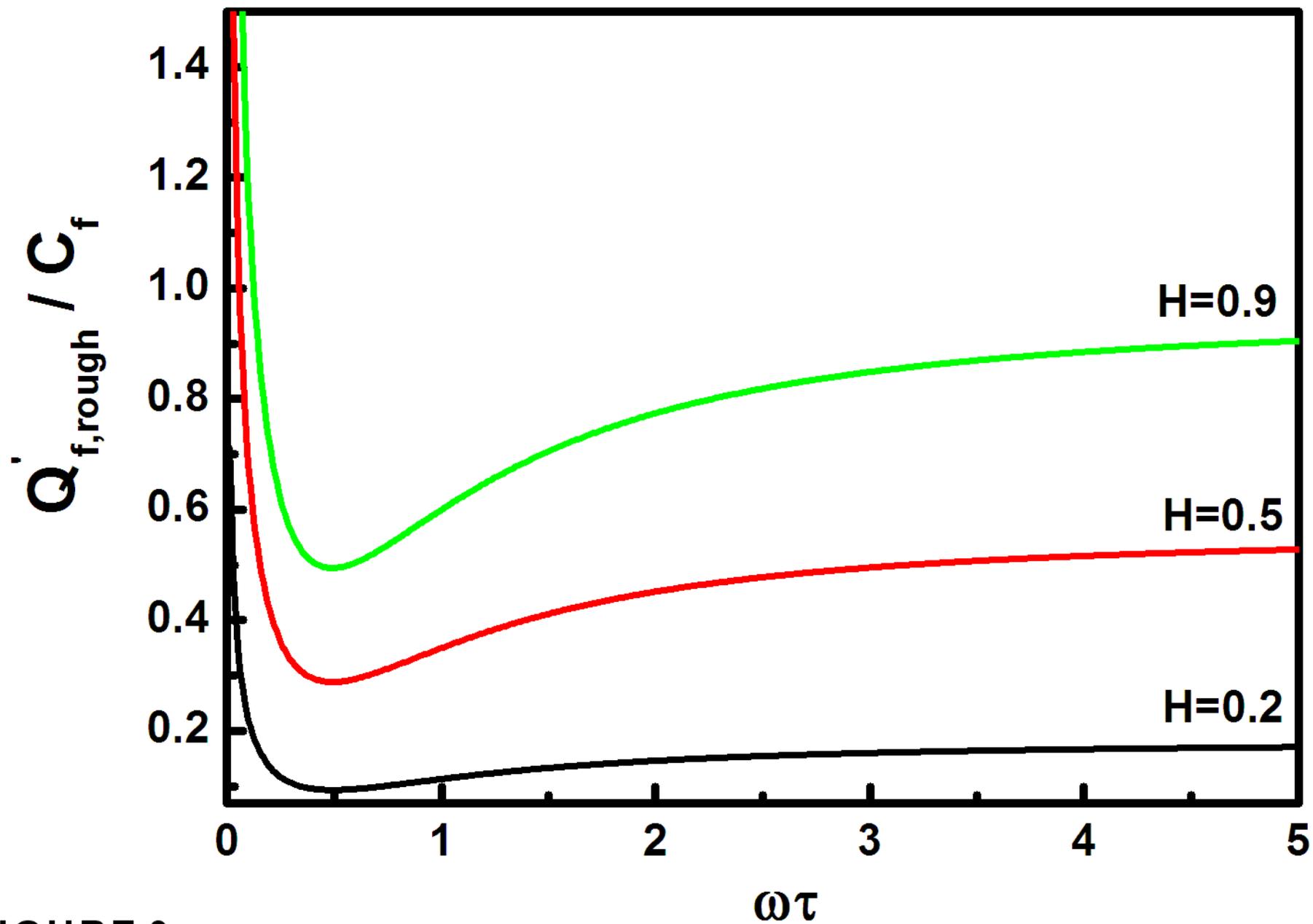

FIGURE 3

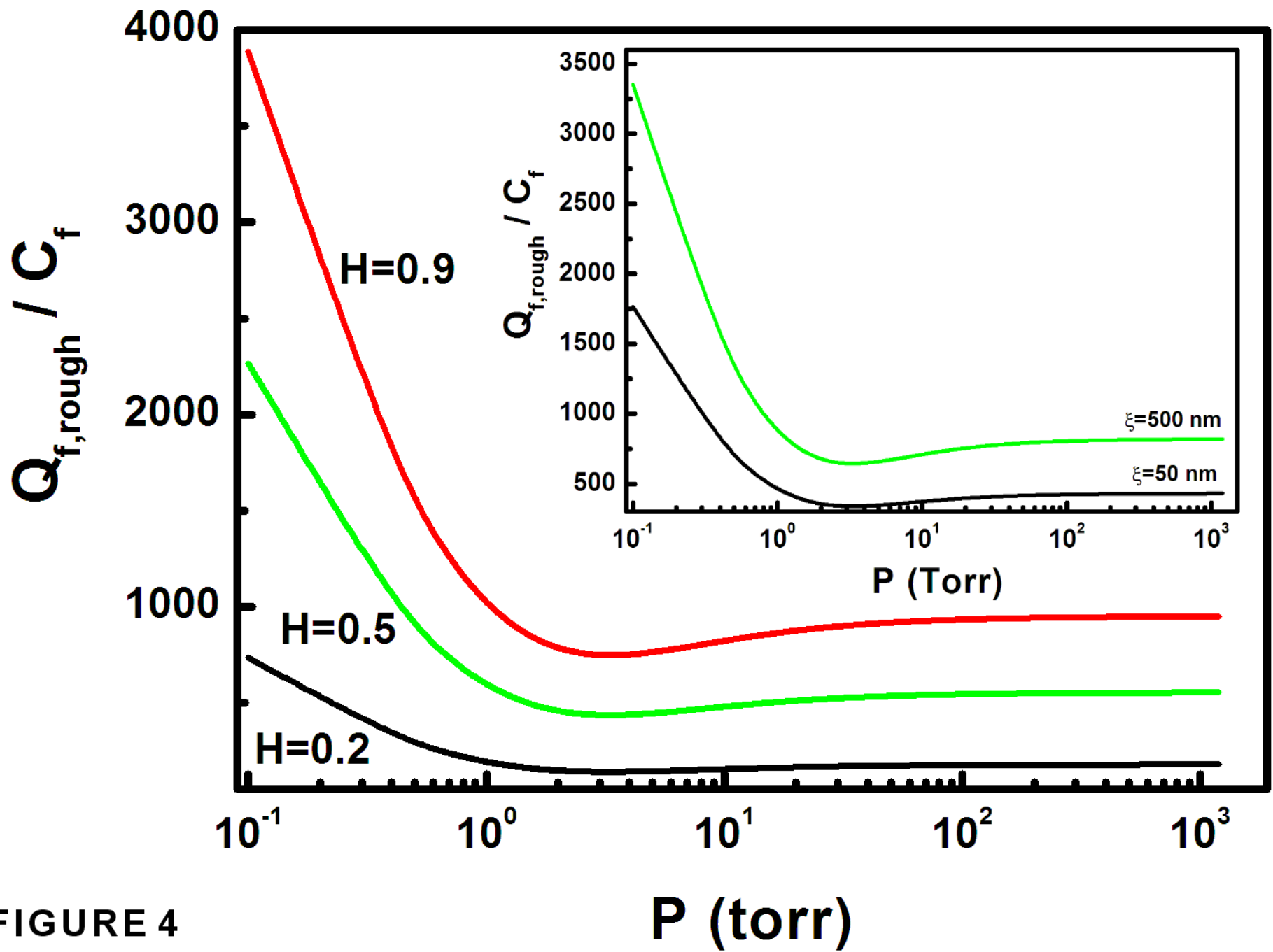

FIGURE 4